\documentclass[aps,amsmath,amssymb,prd,showpacs,floatfix,preprint,superscriptaddress,nofootinbib,12pt]{article}
\usepackage{jheppub}
\usepackage{bm}
\usepackage{ulem}

\usepackage{axodraw4j}
\usepackage{pstricks}
\usepackage{color}

\allowdisplaybreaks[3]

\begin{document}

\title{Critical long-range vector model in the UV}

\author[]{Soumangsu Chakraborty,}
\author[]{Mikhail Goykhman}
\affiliation[]{Department of Theoretical Physics, Tata Institute of Fundamental Research,
Homi Bhabha Road, Mumbai 400005, India}
\affiliation[]{The Racah Institute of Physics, The Hebrew University of Jerusalem, \\ Jerusalem 91904, Israel}
\emailAdd{soumangsuchakraborty@gmail.com}
\emailAdd{michael.goykhman@mail.huji.ac.il}

\abstract{
We study interacting critical UV regime of the long-range $O(N)$ vector model with quartic coupling.
Analyzing CFT data within the scope of $\epsilon$- and $1/N$-expansion,
we collect evidence for the  equivalence of this model and
the critical IR limit of the cubic model coupled to a generalized free field $O(N)$ vector multiplet.
}

\maketitle

\section{Introduction and summary}
\label{sec:intro}

Spin systems with the long-range order, and their continuous
version given by the long-range vector models,
have a long history, starting with the earlier works of \cite{Dyson:1968up,Fisher:1972zz,Kosterlitz:1976,Aizenman1988}.
Such models are characterized by the positive-valued parameter $s$ that controls
the power-law decay of the long-range spin interaction, or a bi-local kinetic
term of the corresponding generalized free field.
Dependence on the parameter $s$ is in particular inherited in the critical
regime of these models, thereby substantially enriching their phase structure.
In this paper we intend to explore the UV critical regime of the $d$-dimensional long-range $O(N)$
vector model with quartic interaction achieved within the domain $s<d/2$.

The long-range $O(N)$ vector model is obtained by perturbing the generalized free field action
for the scalar multiplet $\phi^i$, $i=1,\dots,N$, by the local
quartic interaction term \cite{Fisher:1972zz},
\begin{equation}
\label{long range v.m.}
S = C(s)\,\int d^dx\int d^dy\,\frac{\phi ^i(x)\phi ^i(y)}{|x-y|^{d+s}} + \frac{g_4}{N}\,\int d^dx\,(\phi^i\phi^i)^2\,.
\end{equation}
Here the bi-local kinetic term prefactor
\begin{equation}
C(s) = \frac{2^{s-1}}{\pi^\frac{d}{2}}\,\frac{\Gamma\left(\frac{d+s}{2}\right)}{\Gamma\left(-\frac{s}{2}\right)}
\end{equation}
was chosen so that the propagator of the field $\phi$ is canonically normalized in momentum space,\footnote{See, e.g.,
\cite{Chai:2021arp} for a recent discussion.}
\begin{equation}
\langle \phi ^i(p)\phi ^j(q)\rangle
=(2\pi)^d\,\delta^{(d)}(p+q)\,\delta^{ij}\,\frac{1}{p^s}\,.
\end{equation}
In what follows, we will tend to skip explicitly keeping track of the $O(N)$
indices and the associated Kronecker symbols where it does not cause a confusion.

The scaling dimension of the field $\phi$ is fixed exactly by the bi-local kinetic
term in the action (\ref{long range v.m.}) to be \cite{Fisher:1972zz}
\begin{equation}
\Delta_\phi = \frac{d-s}{2}\,.
\end{equation}
Unitarity bound on $\Delta_\phi$ then restricts us to consider $s\leq 2$.
The quartic coupling constant $g_4$ in the action (\ref{long range v.m.}) is 
relevant when $s>d/2$. In this case the theory flows from the UV fixed
point $g_4 = 0$ to a non-trivial critical regime in the IR. While this theory
does not possess a local stress-energy tensor, a comprehensive argument has been
provided to support the claim that its critical regime is described by an interacting CFT \cite{Brydges:2002wq,Abdesselam:2006qg,Slade:2016yer,Benedetti:2020rrq,Paulos:2015jfa,Behan:2017dwr,Gubser:2017vgc,Behan:2017emf,Behan:2018hfx,Giombi:2019enr,Brezin2014,Chai:2021arp}.
The bounds on $s$ 
impose the condition $d<4$ on the allowed space-time dimension,
that accommodates a non-trivial critical regime in the IR.
When $s=d/2+\epsilon$,
one can study the IR fixed point perturbatively in $\epsilon$ \cite{Fisher:1972zz}. 

Besides being restricted from above by the unitarity bound $s\leq 2$,
one can see that for large enough $s$ the long-range model (\ref{long range v.m.})
will transition to a short-range regime \cite{Sak:1973,Sak:1977}.
In other words, the long-range IR CFT is found in the domain $d/2<s<s_\star$.
The upper bound $s_\star$ of the long-range domain can be determined
by imposing the condition of relevance of the bi-local kinetic term.
Indeed, consider the short-range $O(N)$ vector model describing dynamics of the field $\hat\phi$,
with the local kinetic term $\frac{1}{2}\int d^dx(\partial\hat\phi)^2$, perturbed by a quartic interaction
$(\hat\phi^2)^2$.
At the IR fixed point one obtains the scaling dimension $[\hat\phi]_{\textrm{IR}} = \frac{d}{2} - 1 +\gamma_{\hat\phi}$.
Evaluating the scaling dimension of the bi-local kinetic term for the field $\hat\phi$,
one arrives at $s-s_\star$, where $s_\star=2-2\gamma_{\hat\phi}$. The bi-local kinetic term therefore
becomes more relevant when $s < s_\star$, while the point $s=s_\star$
defines the long-range--short-range crossover. CFT data is expected to be smooth across
the crossover \cite{Sak:1973,Sak:1977}.

The nature of the short-range CFT to the right of the crossover point, $s>s_\star$, was
elucidated in \cite{Behan:2017dwr,Behan:2017emf}. It is given by the critical
short-range $O(N)$ vector model plus a decoupled
generalized free field $\chi$ of dimension $\Delta_\chi = (d+s)/2$. On the other hand, when $d/2<s<s_\star$,
a coupling $\lambda\hat\phi^i\chi^i$ triggers an RG flow, that terminates at the long-range CFT
describing the IR critical regime of the model (\ref{long range v.m.}) \cite{Behan:2017dwr,Behan:2017emf}.
The RG flow triggered by the coupling $\lambda$ can be studied perturbatively in the vicinity of $s_\star$, complementing
a perturbative study near the other edge, $s=d/2$, of the long-range window, $d/2<s<s_\star$
\cite{Behan:2017dwr,Behan:2017emf}.\\

The CFT data of the model (\ref{long range v.m.}) was recently extensively
studied in \cite{Chai:2021arp} in the large $N$ limit up to the next-to-leading
order in $1/N$ expansion, and for general values of $s$ and $d$. To this end, the Hubbard-Stratonovich
formalism was used, that allows for the model to be formulated as
\begin{equation}
\label{long range v.m. in HS}
S = C(s)\,\int d^dx\int d^dy\,\frac{\phi ^i(x)\phi ^i(y)}{|x-y|^{d+s}} +
\int d^dx\,\left(-\frac{1}{4g_4}\,\sigma^2 + \frac{1}{\sqrt{N}}\,\sigma\phi^2\right)\,.
\end{equation}
At the fixed point the scaling dimension of the Hubbard-Stratonovich field $\sigma$ is given by \cite{Gubser:2017vgc,Giombi:2019enr,Chai:2021arp}
\begin{align}
\label{[sigma] in v.m.}
[\sigma] _{\textrm{crit}} &= s + \tilde\gamma_\sigma\,,\\
\tilde\gamma_\sigma &{=} -\frac{1}{N}\frac{4 \Gamma \left(\frac{s}{2}\right)^2 \Gamma (d{-}s)}
{\Gamma \left(\frac{d}{2}\right) \Gamma (s) \Gamma
\left(d{-}\frac{3 s}{2}\right) \Gamma \left(\frac{d{-}s}{2}\right)^3 \Gamma \left(s{-}\frac{d}{2}\right)^2}
\left(\Gamma (s) \Gamma \left(d{-}\frac{3 s}{2}\right) \Gamma \left(\frac{d{-}s}{2}\right)
\Gamma \left(s{-}\frac{d}{2}\right)\right.\notag\\
&-\left. 2 \Gamma \left(\frac{s}{2}\right) \Gamma \left(\frac{d}{2}-s\right)
\Gamma (d-s) \Gamma \left(\frac{3 s}{2}-\frac{d}{2}\right)\right)+{\cal O}\left(\frac{1}{N^2}\right)\,.\label{gamma sigma in v.m.}
\end{align}
For $s>d/2$, in the large $N$ limit, the $\sigma^2$ term in (\ref{long range v.m. in HS}) is then irrelevant in the IR,
while for $s<d/2$
it is irrelevant in the UV, and has no influence on the corresponding CFT.
Notice that near $s=d/3$ we have
\begin{equation}
\tilde\gamma_\sigma = \frac{1}{N}\,\frac{\kappa}{s-\frac{d}{3}}+{\cal O}\left(\left(s-\frac{d}{3}\right)^0\right)\,,
\end{equation}
where we defined the positive-valued coefficient
\begin{equation}
\kappa = \frac{16 \Gamma \left(\frac{d}{6}\right)^4 \Gamma \left(\frac{2 d}{3}\right)^2}{3 \Gamma \left(-\frac{d}{6}\right)^2 \Gamma \left(\frac{d}{3}\right)^4 \Gamma \left(\frac{d}{2}\right)^2}\,.
\end{equation}
Therefore for $s<d/3$ one can obtain arbitrarily large 
negative-valued $\tilde\gamma_\sigma$, resulting in the dimension $[\sigma]_{\textrm{crit}}$ dipping
below the unitarity bound. Such a regime needs to be excluded.

To summarize, the model (\ref{long range v.m. in HS}) achieves a non-trivial critical regime in the
IR for
\begin{equation}
d/2<s<s_\star\,,\qquad 2<d<4\,,
\end{equation}
and in the UV for
\begin{equation}
\label{d/3 bound on s}
d/3 < s <\min(d/2, s_\star)\,, \qquad 2<d<6\,.
\end{equation}
Notice that since at the leading order in $1/N$ we have $s_\star = 2+{\cal O}(1/N)$,
then in the large $N$ limit the upper bound in (\ref{d/3 bound on s}) is determined by whether we
are working in low ($2<d<4$) or higher ($4<d<6$) dimensions.

The CFT data derived in \cite{Chai:2021arp} includes the ratio of the large-$N$ three-point functions
amplitudes (OPE coefficients)\footnote{The tag `normalized' means that
 the propagators of the fields $\phi$ and $\sigma$ in position space have been normalized to unity.}
\begin{equation}
\label{sigma sigma sigma over phi phi sigma in v.m.}
\frac{\langle \sigma\sigma\sigma\rangle}{\langle \sigma\phi\phi\rangle} \Bigg|_{\textrm{normalized}}{=} 
\frac{2^{d{-}2 s} (d{-}2 s) \sin \left(\frac{1}{2} \pi  (d{-}2 s)\right) \Gamma \left(\frac{d}{2}{-}s\right)^2
\Gamma \left(\frac{1}{2} (d{-}s{+}1)\right) \Gamma \left(\frac{1}{2} (3 s{-}d)\right)}
{\pi  \Gamma \left(\frac{s{+}1}{2}\right) \Gamma \left(d{-}\frac{3 s}{2}\right)}\,,
\end{equation}
as well as the ratio of the next-to-leading to leading (in $1/N$
expansion) OPE coefficient $\langle \sigma\phi\phi\rangle$
\begin{equation}
\label{delta C phi phi sigma in v.m.}
\delta C_{\phi\phi\sigma}\Bigg|_{\textrm{normalized}}=
\delta Z_{\phi\phi\sigma} +\delta {\cal U}_{\phi\phi\sigma} +A_\phi+\frac{A_\sigma}{2}\,,
\end{equation}
where we refer the reader to \cite{Chai:2021arp} for details.\\

Our discussion above was centered around the goal to obtain
a non-trivial fixed point in the IR. The corresponding arguments parallel those
of the ordinary short-range (local) $O(N)$ vector model, that
features a non-trivial IR criticality when $2<d<4$. When $d=4-\epsilon$,
the corresponding critical regime of the short-range vector model admits a perturbative 
Wilson-Fisher description \cite{Wilson:1971dc}. Pushing to higher dimensions, when $d=4+\epsilon$,
the WF approach can be analytically continued, indicating existence of a fixed
point in the UV rather than in the IR. This fixed point is achieved at negative values
of the quartic coupling constant, $g_4$, and therefore the model is unstable. In fact, an interacting UV fixed point
of the short-range $O(N)$ vector model is believed
to exist in the entire range $4<d<6$, where the upper bound $d=6$ is due to
the unitarity constraint on the scaling dimension of the Hubbard-Stratonovich field \cite{Parisi:1975im}.
Given the problematic nature of this UV fixed point, it is desirable to get a better understanding
of the corresponding CFT. This motivated \cite{Fei:2014yja} to argue for a critical duality 
between the UV fixed point of the higher-dimensional $O(N)$ vector model and the IR fixed point of the certain cubic model.
Such a duality (critical universality or, more specifically, a UV completion)
has been successfully tested up to the fourth order in perturbation theory
in $d=6-\epsilon$ dimensions, and in fact was argued to hold true in the entire range $4<d<6$ \cite{Fei:2014xta,Gracey:2015tta}.\footnote{Formally one can consider the $O(N)$ vector model in the non-unitary domain $d>6$,
where yet another model was argued in \cite{Gracey:2015xmw} to be critically
equivalent to the $O(N)$ vector model.}

Similarly to the case of the short-range $O(N)$ vector model, described in the previous paragraph,
the non-trivial criticality region $s>d/2$ of the long-range vector model can be extended to $s<d/2$
if we look for an interacting fixed point in the UV rather than in the IR. Indeed, when $s<d/2$,
the quartic interaction in the long-range model (\ref{long range v.m.})
is irrelevant, indicating an RG flow to an interacting  regime in the UV.
Together with the unitarity bound $s< 2$ and the long-range bound $s<s_\star$ we obtain $s<\min\left(d/2,s_\star\right)$.\footnote{Recall that $\gamma_{\hat\phi} > 0$  \cite{Fei:2014xta,Petkou:1995vu,Petkou:1994ad,Vasiliev:1981yc,Vasiliev:1981dg} and therefore $s_\star < 2$.}
Analogously to the local short-range case, the instability issue
is found yet again at the corresponding UV fixed point, as can be seen by a perturbative
study at $s=d/2-\epsilon$, rendering a negative fixed point value of the quartic coupling constant.
The question is then whether one can construct a UV completion of the corresponding CFT in the spirit
of \cite{Fei:2014yja}. This is the problem that we address in the current paper.\\

We propose that the UV fixed point of the model (\ref{long range v.m.}) within the range $0<s<d/2$ is 
identical (critically dual) to the IR fixed point of the following model\footnote{Since from now one we will
be mostly discussing the model (\ref{starting bilocal phi action}), we are going
to recycle the symbols $S$, $\phi$, $\sigma$.}
\begin{align}
\label{starting bilocal phi action}
S = C(s)\,\int d^dx\int d^dy\,\frac{\phi(x)\phi(y)}{|x-y|^{d+s}}
+\int d^dx\,\left(\frac{1}{2}(\partial\sigma)^2 +\frac{1}{2}g\sigma\phi^2
+\frac{1}{6}h\sigma^3\right)\,,
\end{align}
describing dynamics of the vector multiplet of generalized free field $\phi^i$, $i=1,\dots,N$,
coupled
to the singlet scalar $\sigma$ with the cubic self-interaction. Scaling dimensions
of these fields are given by
\begin{equation}
\label{UV dimensions in the cubic model}
[\phi] = \Delta_\phi = \frac{d-s}{2}\,,\qquad [\sigma]_{\textrm{UV}} =  \Delta_\sigma = \frac{d-2}{2}\,.
\end{equation}
Unitarity then demands $s<2$, which leads to $s<\min\left(d/2,2\right)$.
In fact, the model (\ref{starting bilocal phi action}) transitions
to the short-range regime when the $s>s_\star$,
that can be determined by calculating scaling dimension of the bi-local
kinetic term for the field $\tilde\phi$ in the critical cubic short-range model.
Scaling dimension of the latter is equal to the scaling
dimension of the field $\hat\phi$ of the short-range $O(N)$
vector model with quartic interaction \cite{Fei:2014yja}.
Therefore at the outset we restrict to the range $0<s<\min\left(d/2, s_\star\right)$.

Similarly to \cite{Fei:2014yja}, we are looking
for an IR stable fixed point on the $(g,h)$ plane.
We will argue for the equivalence of the corresponding critical theory
and the UV fixed point of the $O(N)$ vector model (\ref{long range v.m.}),
by calculating and matching CFT data,
such as scaling dimensions and OPE coefficients.\footnote{Another example
of a UV completion of non-local vector model was discussed in \cite{Giombi:2019enr}, that considered the case of
$s=1$, corresponding to a boundary CFT.}

Establishing the critical duality between the models (\ref{long range v.m.}), (\ref{starting bilocal phi action})
begins with matching of the d.o.f. To that end, we use the Hubbard-Stratonovich representation (\ref{long range v.m. in HS})
of the long-range vector model.
The fields $\phi^i$ and $\sigma$ are accordingly matched in (\ref{long range v.m. in HS}), (\ref{starting bilocal phi action}).
By construction, the scaling dimension $[\phi] =\Delta_\phi = (d-s)/2$ is exact
in interacting theory in both of these models. Below in this paper we will also match the CFT data
of the cubic model (\ref{starting bilocal phi action}) with its counterparts
(\ref{[sigma] in v.m.}), (\ref{gamma sigma in v.m.}), (\ref{sigma sigma sigma over phi phi sigma in v.m.}),
(\ref{delta C phi phi sigma in v.m.}) in the long-range vector model (\ref{long range v.m. in HS}).

We will study the model (\ref{starting bilocal phi action})
in a perturbative regime in the couplings $g$, $h$, by choosing the specific
values of the parameters $d$, $s$.
Mass dimensions of the coupling constants $g$, $h$ in the action (\ref{starting bilocal phi action})
are given by
\begin{equation}
\label{g and h scaling dimensions}
[g]_{\textrm{UV}} = s+1-\frac{d}{2} = \epsilon_2\,,\qquad [h]_{\textrm{UV}} = 3-\frac{d}{2} = \frac{\epsilon_1}{2}\,,
\end{equation}
where we defined
\begin{equation}
\label{d and s in terms of epsilon12}
d = 6-\epsilon_1\,,\qquad s=\frac{d}{2}-1 +\epsilon_2=2-\frac{\epsilon_1}{2}+\epsilon_2\,.
\end{equation}
Therefore, relevance of the couplings in the UV imposes the constraints
\begin{equation}
d\leq 6\,,\qquad  s\geq d/2-1\,,
\label{bounds on d and s}
\end{equation}
that lead to $\epsilon_{1,2} \geq 0$. Notice that the constraint on $s$
in (\ref{bounds on d and s}) selects a sub-region $d/2-1\leq s <\min(d/2, s_\star)$
within the domain $s<\min\left(d/2, s_\star\right)$, where the long-range $O(N)$
vector model flows to a non-trivial critical regime in the UV.

One can then perform Wilson-Fisher perturbative expansion in $\epsilon_{1,2}$.
We will be working at the linear order ${\cal O}(\epsilon_{1,2})$.
It is convenient to denote
\begin{equation}
\label{alpha definition}
\alpha = \frac{\epsilon_2}{\epsilon_1} < \frac{1}{2}+{\cal O}\left(\frac{1}{N}\right)\,,
\end{equation}
where the last inequality follows from the bound $s <\min(d/2, s_\star) =  s_\star=2+{\cal O}(1/N)$,
taking into account $d=6-\epsilon_1$ and $\gamma_{\tilde\phi} = {\cal O}(1/N)$.
 Our results will be exact to all orders in $1/N$.
Subsequently performing algebraic expansion to the next-to-leading order in $1/N$ one can compare the CFT data
that we derive in this paper
with its counterpart in the long-range vector model. The latter (having been
derived in $1/N$ expansion but for general $d$ and $s$) is in turn to be 
expanded to a linear order in $\epsilon_{1,2}$.\\

While specifics of the match of CFT data on both sides of the proposed critical duality
will be carefully detailed below in this paper, we can briefly outline some important
ingredients that already supply a promising indication that the duality is going to hold
(at least within the $\epsilon$-expansion). 
The fixed point values for the couplings $g$, $h$, determined at the leading order in $\epsilon_{1,2}$,
and subsequently expanded in $1/N$, have the form
\begin{align}
\label{1/N expansion of couplings}
g &=g_0\,\left(1+\frac{g_1}{N} + {\cal O}\left(\frac{1}{N^2}\right)\right)\,,\\
h &=h_0\,\left(1+\frac{h_1}{N} + {\cal O}\left(\frac{1}{N^2}\right)\right)\,,
\end{align}
where
\begin{equation}
g_0\,, \; h_0 = {\cal O}\left(\sqrt{\frac{\epsilon_1}{N}}\right)\,.
\end{equation}
One can see that the Callan-Symanzik equation for the $\langle \phi\phi\sigma\rangle$
three-point function then implies (here we take into account that $\gamma_\phi=0$)\footnote{
At the same time, the Callan-Symanzik equation for the $\langle \sigma\sigma\sigma\rangle$ three-point function
at the leading order ${\cal O}(1/N^{1/2})$ includes a contribution from the counterterm $\delta h$.
We consistently solve the Callan-Symanzik equations for both $\langle \phi\phi\sigma\rangle$ and
$\langle \sigma\sigma\sigma\rangle$  in the main text.}
\begin{equation}
\label{CS derivation of gamma sigma}
\beta_g = -\epsilon_2 \,g + g\,\gamma_\sigma + {\cal O}\left(\frac{1}{N^{3/2}}\right)\,,
\end{equation}
and therefore at the non-trivial IR fixed point we obtain anomalous dimension
\begin{equation}
\label{gamma sigma from CS}
\gamma_\sigma = \epsilon _ 2 + {\cal O}\left(\frac{1}{N}\right)\,.
\end{equation}
Consequently the IR scaling dimension of the field $\sigma$ is given by
\begin{equation}
\label{total IR dimension of sigma}
[\sigma]_{\textrm{IR}} = \frac{d-2}{2} + \gamma_\sigma = s + {\cal O}\left(\frac{1}{N}\right)\,.
\end{equation}
Notice that unlike the case (\ref{gamma sigma in v.m.})
of the long-range $O(N)$ vector model, the anomalous dimension $\gamma_\sigma$
has ${\cal O}(1/N^0)$ component. Combined with the UV dimension (\ref{UV dimensions in the cubic model})
it results in the large-$N$ expression (\ref{total IR dimension of sigma}), that agrees with its counterpart
in the critical long-range vector model (\ref{[sigma] in v.m.}).

Furthermore, we obtain the IR scaling dimension of the cubic coupling $[h]_{\textrm{IR}} = d-3s+{\cal O}\left(\frac{1}{N}\right)$.
The cubic term, whose presence in the model (\ref{starting bilocal phi action})
naively distinguishes it 
from the critical $O(N)$ vector model (\ref{long range v.m. in HS}), is then irrelevant in the IR,
provided that $s>d/3$. This bound is compatible with the unitarity bound $s\leq 2$,
if we demand that $d\leq 6$. Such a restriction on space-time dimension is also compatible
with our assumptions (\ref{bounds on d and s}), that also makes
the $s>d/3$ constraint stronger than $s\geq d/2-1$ constraint in (\ref{bounds on d and s}).
This means that the region that the model (\ref{starting bilocal phi action}) allows us to study is\footnote{
Similarly, one can see that the local kinetic term for $\sigma$ in
(\ref{starting bilocal phi action}) is irrelevant in the IR when $s> d/2-1$, in agreement with
(\ref{bounds on d and s}).}
\begin{equation}
\label{d/3 bound on s rep}
d/3<s<\min(d/2, s_\star)\,,
\end{equation}
that agrees with (\ref{d/3 bound on s}).
In terms of $\epsilon_{1,2}$ it gives $\alpha > 1/6$,\footnote{Indeed, using (\ref{d and s in terms of epsilon12}) we obtain
$s-d/3=\epsilon_1\,(\alpha-1/6)>0$, landing $s$ within the range (\ref{d/3 bound on s rep}).}
that in combination with (\ref{alpha definition}) defines the domain
\begin{equation}
\label{bounds on alpha}
\frac{1}{6} < \alpha < \frac{1}{2}\,.
\end{equation}
\\

The rest of this paper is organized as follows. In section~\ref{sec:anomalous dimensions and beta functions}
we will carry out renormalization of the model (\ref{starting bilocal phi action}) to linear order in perturbation theory,
deriving anomalous dimension of $\sigma$ and the beta functions for the couplings $g$, $h$.
We solve for the fixed points of this model in section~\ref{sec:fixed points}, and determine the IR stable
fixed point on the $(g,h)$ plane.
In section~\ref{sec:CFT data} we analyze the CFT data at the IR stable fixed point, and demonstrate that it matches
the known CFT data at the UV fixed point of the $O(N)$ vector model.
We discuss our results and possible future directions in section~\ref{sec:discussion}.
We assemble some well-known identities, useful for calculations
in conformal perturbation theory in position space, in appendix~\ref{app_a}.

\section{Anomalous dimensions and beta functions}
\label{sec:anomalous dimensions and beta functions}

We will perform our calculations in the model (\ref{starting bilocal phi action})
perturbatively in the couplings $g$, $h$ near the free UV fixed point, working at
the linear order in $\epsilon_{1,2}$.
Free propagators of the fields $\phi$, $\sigma$ in position space are given by
\begin{equation}
\label{leading order propagators}
\langle \phi(x)\phi(0)\rangle = \frac{C_\phi}{|x|^{2\Delta_\phi}}\,,\qquad
\langle \sigma(x)\sigma(0)\rangle = \frac{C_\sigma}{|x|^{2\Delta_\sigma}}\,,
\end{equation}
where the propagator amplitudes are given by
\begin{equation}
C_\phi = \frac{1}{2^s\pi^\frac{d}{2}}\,\frac{\Gamma\left(\frac{d-s}{2}\right)}{\Gamma\left(\frac{s}{2}\right)}\,,\qquad
C_\sigma = \frac{\Gamma\left(\frac{d-2}{2}\right)}{4\pi^\frac{d}{2}}\,.
\end{equation}
Interactions will introduce corrections to the propagator amplitudes, that are
not essential for our goals.
In fact, we will only be interested in the singular parts of the correlation functions
that we compute.
As we mentioned in the previous section, the scaling dimension $\Delta_\phi$ of $\phi$
will remain exact, while the scaling dimension $\Delta_\sigma$ of $\sigma$ will receive an anomalous contribution,
that we intend to calculate.

We are going to employ perturbation theory in position space.
The corresponding Feynman rules are given by
\begin{center}
  \begin{picture}(350,67) (24,-17)
    \SetWidth{1.0}
    \SetColor{Black}
    \Line[](34,34)(98,34)
    \Vertex(34,34){2}
    \Vertex(98,34){2}
    \Text(106,23)[lb]{\scalebox{1.01}{$=\frac{C_\phi}{|x|^{2\Delta_\phi}}$}}
    \Text(106,-15)[lb]{\scalebox{1.01}{$=\frac{C_\sigma}{|x|^{2\Delta_\sigma}}$}}
    \Text(98,26)[lb]{\scalebox{0.81}{$x$}}
    \Text(24,26)[lb]{\scalebox{0.81}{$0$}}
    \Text(61,36)[lb]{\scalebox{0.81}{$2\Delta_\phi$}}
    \Line[](32,-6)(96,-6)
    \Text(21,-14)[lb]{\scalebox{0.81}{$0$}}
    \Text(98,-14)[lb]{\scalebox{0.81}{$x$}}
    \Text(60,-4)[lb]{\scalebox{0.81}{$2\Delta_\sigma$}}
    \Vertex(32,-6){2}
    \Vertex(96,-6){2}
    \Line[](185,34)(230,14)
    \Line[](230,14)(185,-6)
    \Line[](230,14)(272,14)
    \Vertex(230,14){4}
    \Text(165,-10)[lb]{\scalebox{0.81}{$\phi,\sigma$}}
    \Text(165,30)[lb]{\scalebox{0.81}{$\phi,\sigma$}}
    \Text(279,12)[lb]{\scalebox{0.81}{$\sigma$}}
    \Text(289,8)[lb]{\scalebox{0.9}{$=- g\mu^{\epsilon_2},- h\mu^{\epsilon_1/2}\,.$}}
  \end{picture}
\end{center}
Here we have also introduced the dimensionless couplings $g$, $h$,
taking into account the scaling dimensions (\ref{g and h scaling dimensions}),
and defining an arbitrary RG scale $\mu$.
(To eliminate clutter, we avoid using new notation for the dimensionless couplings.)
A propagator line with a generic exponent will be assumed to have a unit amplitude
\begin{center}
  \begin{picture}(300,7) (24,32)
    \SetWidth{1.0}
    \SetColor{Black}
    \Line[](34,34)(98,34)
    \Vertex(34,34){2}
    \Vertex(98,34){2}
    \Text(63,39)[lb]{\scalebox{0.81}{$2a$}}
    \Text(98,26)[lb]{\scalebox{0.81}{$x$}}
    \Text(24,26)[lb]{\scalebox{0.81}{$0$}}
    \Text(106,25)[lb]{\scalebox{1.01}{$=\frac{1}{|x|^{2a}}\,.$}}
    \end{picture}
\end{center}
Diagrammatic calculations in position space can be performed using various well-known identities that we
collected in appendix~\ref{app_a}.\\

The linear order, correction to the $\langle\phi\phi\rangle$
propagator is determined by the diagram
\begin{center}
  \begin{picture}(396,60) (48,10)
    \SetWidth{1.0}
    \SetColor{Black}
    \Line[](168,37)(224,37)
    \Arc[](252,37)(25.239,146,506)
    \Line[](280,37)(336,37)
    \Vertex(168,37){2}
    \Vertex(227,37){4}
    \Vertex(277,37){4}
    \Vertex(336,37){2}
    \Text(189,40)[lb]{\scalebox{0.8}{$2\Delta_\phi$}}
    \Text(245,65)[lb]{\scalebox{0.8}{$2\Delta_\phi$}}
    \Text(247,0)[lb]{\scalebox{0.8}{$2\Delta_\sigma$}}
    \Text(301,40)[lb]{\scalebox{0.8}{$2\Delta_\phi$}}
  \end{picture}
\end{center}
Combining this with the leading order propagator (\ref{leading order propagators}), we obtain
\begin{align}
\label{corrected phi phi}
\langle \phi(x)\phi(0)\rangle &\supset
\frac{C_\phi}{|x|^{d-s}}\,\left( 1 + g^2\,(\mu |x|)^{2\epsilon_2}\, C_\phi^2 \, C_\sigma\, V^{(\phi\phi)}\right)\,,\\
V^{(\phi\phi)} &= U\left(\frac{d-s}{2},d-1-\frac{s}{2},1+s-\frac{d}{2}\right)
U\left(\frac{d-s}{2},d-1-s,1+\frac{3s}{2}-\frac{d}{2}\right)\,,\notag
\end{align}
where we took advantage of the simple behavior of loops in position 
space, discussed in appendix~\ref{app_a}, and used the propagator merging relation (\ref{prop merging})
and the function $U(a,b,c)$, given by (\ref{U def}).
Expanding the second term in the parenthesis in (\ref{corrected phi phi}) in $\epsilon_2$,
we obtain a finite contribution, implying $\gamma_\phi = 0$, in agreement with our expectation that scaling dimension
of $\phi$ does not receive anomalous contributions.\\

Similarly, we have the following contributions to the $\langle\sigma\sigma\rangle$
propagator at the one-loop order in perturbation theory:
\begin{center}
  \begin{picture}(200,60) (248,10)
    \SetWidth{1.0}
    \SetColor{Black}
    \Line[](168,37)(224,37)
    \Arc[](252,37)(25.239,146,506)
    \Line[](280,37)(336,37)
    \Vertex(168,37){2}
    \Vertex(227,37){4}
    \Vertex(277,37){4}
    \Vertex(336,37){2}
    \Text(189,40)[lb]{\scalebox{0.8}{$2\Delta_\sigma$}}
    \Text(245,65)[lb]{\scalebox{0.8}{$2\Delta_\phi$}}
    \Text(247,0)[lb]{\scalebox{0.8}{$2\Delta_\phi$}}
    \Text(301,40)[lb]{\scalebox{0.8}{$2\Delta_\sigma$}}
    \Text(350,34)[lb]{\scalebox{0.8}{$+$}}
    \Line[](368,37)(424,37)
    \Arc[](452,37)(25.239,146,506)
    \Line[](480,37)(536,37)
    \Vertex(368,37){2}
    \Vertex(427,37){4}
    \Vertex(477,37){4}
    \Vertex(536,37){2}
    \Text(389,40)[lb]{\scalebox{0.8}{$2\Delta_\sigma$}}
    \Text(445,65)[lb]{\scalebox{0.8}{$2\Delta_\sigma$}}
    \Text(447,0)[lb]{\scalebox{0.8}{$2\Delta_\sigma$}}
    \Text(501,40)[lb]{\scalebox{0.8}{$2\Delta_\sigma$}}
  \end{picture}
\end{center}
Together with the leading order propagator (\ref{leading order propagators}) we get
\begin{align}
\label{sigma sigma 1loop correction}
\langle \sigma(x)\sigma(0)\rangle
\supset \frac{C_\sigma}{|x|^{d-2}}\,\left( 1 +\frac{1}{2} \,C_\sigma\,C_\phi^2\, g^2\, (\mu |x|)^{2\epsilon_2}\, V^{(\sigma\sigma)}_1
+\frac{1}{2} \,C_\sigma^3\,h^2\,(\mu |x|)^{\epsilon_1}\, V^{(\sigma\sigma)}_2\right)\,,
\end{align}
where we denoted
\begin{align}
V^{(\sigma\sigma)}_1 &= U\left(\frac{d}{2}-1,d-s,s+1-\frac{d}{2}\right)U\left(\frac{d}{2}-1,d-s-1,s+2-\frac{d}{2}\right)\,,\\
V^{(\sigma\sigma)}_2 &= U\left(\frac{d}{2}-1,d-2,3-\frac{d}{2}\right)U\left(\frac{d}{2}-1,d-3,4-\frac{d}{2}\right)\,.
\end{align}
Expanding the relative corrections in the parenthesis of (\ref{sigma sigma 1loop correction}) around
$\epsilon_2= 0$ and $\epsilon_1= 0$ respectively,
we obtain
\begin{equation}
\label{gamma sigma}
\begin{aligned}
\langle \sigma(x)\sigma(0)\rangle &\supset \frac{C_\sigma}{|x|^{d-2}}\,\left( 1 - 2\gamma_\sigma \,\log(\mu |x|)\right)\,,\\
\gamma_\sigma &= \frac{h^2 + N \,g^2}{12(4\pi)^3} + {\cal O}(\epsilon_{1,2}^{2})\,,
\end{aligned}
\end{equation}
where we absorbed the pure divergence into the counterterm due to the wave-function renormalization
of $\sigma$, that we have kept implicit.\\

We now proceed to the perturbative calculation of the three-point function $\langle \phi\phi\sigma\rangle$.
At the leading order, it is given by a tree-level diagram, defined by the Feynman rule for the vertex corresponding to
the coupling constant $g$. To get the three-point function $\langle \phi\phi\sigma\rangle$,
we need to attach to this vertex two $\phi$ and one $\sigma$ propagator legs, and integrate over the location
of the vertex insertion point. Using the uniqueness relation (\ref{uniqueness}), we obtain
\begin{equation}
\label{leading order phi phi sigma}
\langle \phi (x_1)\phi (x_2)\sigma (x_3)\rangle
=\frac{-g\,\mu^{\epsilon_2}\,C_\phi^2\, C_\sigma \, U(2,2,2)|_{d=6}}{(|x_{12}||x_{13}||x_{23}|)^{2}}\,.
\end{equation}

At one-loop order, corrections are given by the following one-loop diagrams
\begin{center}
  \begin{picture}(163,114) (107,-10)
    \SetWidth{1.0}
    \SetColor{Black}
    \Line[](81,95)(81,51)
    \Line[](50,9)(110,10)
    \Line[](9,-7)(51,9)
    \Line[](151,-7)(110,10)
    \Line[](48,8)(81,51)
    \Line[](81,51)(110,10)
    \Vertex(50,9){4}
    \Vertex(81,51){4}
    \Vertex(110,10){4}
    \Vertex(10,-7){2}
    \Vertex(150,-7){2}
    \Vertex(81,95){2}
    \Text(87,70)[lb]{\scalebox{0.8}{$2\Delta_\sigma$}}
    \Text(45,30)[lb]{\scalebox{0.8}{$2\Delta_\phi$}}
    \Text(100,30)[lb]{\scalebox{0.8}{$2\Delta_\phi$}}
    \Text(74,-4)[lb]{\scalebox{0.8}{$2\Delta_\sigma$}}
    \Text(17,5)[lb]{\scalebox{0.8}{$2\Delta_\phi$}}
    \Text(130,5)[lb]{\scalebox{0.8}{$2\Delta_\phi$}}
    \Text(180,30)[lb]{\scalebox{0.8}{$+$}}
    \Line[](281,95)(281,51)
    \Line[](250,9)(310,10)
    \Line[](209,-7)(251,9)
    \Line[](351,-7)(310,10)
    \Line[](248,8)(281,51)
    \Line[](281,51)(310,10)
    \Vertex(250,9){4}
    \Vertex(281,51){4}
    \Vertex(310,10){4}
    \Vertex(210,-7){2}
    \Vertex(350,-7){2}
    \Vertex(281,95){2}
    \Text(287,70)[lb]{\scalebox{0.8}{$2\Delta_\sigma$}}
    \Text(245,30)[lb]{\scalebox{0.8}{$2\Delta_\sigma$}}
    \Text(300,30)[lb]{\scalebox{0.8}{$2\Delta_\sigma$}}
    \Text(274,-4)[lb]{\scalebox{0.8}{$2\Delta_\phi$}}
    \Text(217,5)[lb]{\scalebox{0.8}{$2\Delta_\phi$}}
    \Text(330,5)[lb]{\scalebox{0.8}{$2\Delta_\phi$}}
  \end{picture}
\end{center}
We are interested in the divergent structure of these diagrams, that can be extracted from integrating
over the three internal vertices of the following graph
\begin{equation}
  \begin{picture}(163,114) (7,-10)
    \SetWidth{1.0}
    \SetColor{Black}
    \Line[](81,95)(81,51)
    \Line[](50,9)(110,10)
    \Line[](9,-7)(51,9)
    \Line[](151,-7)(110,10)
    \Line[](48,8)(81,51)
    \Line[](81,51)(110,10)
    \Vertex(50,9){4}
    \Vertex(81,51){4}
    \Vertex(110,10){4}
    \Vertex(10,-7){2}
    \Vertex(150,-7){2}
    \Vertex(81,95){2}
    \Text(87,70)[lb]{\scalebox{0.8}{$4$}}
    \Text(55,30)[lb]{\scalebox{0.8}{$4$}}
    \Text(100,30)[lb]{\scalebox{0.8}{$4$}}
    \Text(78,0)[lb]{\scalebox{0.8}{$4$}}
    \Text(22,5)[lb]{\scalebox{0.8}{$4$}}
    \Text(130,5)[lb]{\scalebox{0.8}{$4$}}
  \end{picture}
  \label{one loop vertex graph}
\end{equation}
After one of the three vertices is integrated over, via the uniqueness relation (\ref{uniqueness}),
producing the factor of $U(2,2,2)|_{d=6}=\pi^3$,
we will encounter a logarithmically divergent integral, that we regulate
using the UV cutoff $\mu_0$,
\begin{equation}
\int \frac{d^6x}{|x|^6} \Bigg|_{|x|>1/\mu_0}= \pi^6\,\log(\mu_0)\,.
\end{equation}
Such divergences are in fact removed with the $\delta g$ and $\delta h$ counterterms (that we keep
mostly implicit in our calculation)
proportional to $-\log(\mu_0/\mu)$, at the expense 
of introducing an arbitrary RG scale $\mu$. The final integral is taken using
the uniqueness relation, producing another factor of  $U(2,2,2)|_{d=6}=\pi^3$.
Assembling everything together, we obtain that the regularized graph (\ref{one loop vertex graph}) is given by
\begin{equation}
V^{(\phi\phi\sigma)} = \pi^{12}\,\log(\mu)\,.
\end{equation}
As a result, we obtain the one-loop corrections to $\langle \phi\phi\sigma\rangle$ given by
\begin{equation}
\label{vertex corrections to phi phi sigma}
\langle \phi (x_1)\phi (x_2)\sigma (x_3)\rangle
\supset\left(-g^3\, C_\phi^4\, C_\sigma^2\, V^{(\phi\phi\sigma)}-g^2 \, h \, C_\phi^3\, C_\sigma^3 \, V^{(\phi\phi\sigma)} 
\right)\,\frac{1}{(|x_{12}||x_{13}||x_{23}|)^{2}}\,.
\end{equation}

Finally, extracting the $\mu$-dependent contribution obtained by dressing the $\sigma$ propagator 
of the tree level diagram (since $\phi$ does not have anomalous dimension,
its propagator does not acquire a $\mu$-dependent dressing), we obtain
\begin{equation}
\label{leading order phi phi sigma with dressed sigma legs}
\langle \phi (x_1)\phi (x_2)\sigma (x_3)\rangle
\supset\frac{2\,g\,\gamma_\sigma\,C_\phi^2\, C_\sigma \, U(2,2,2)|_{d=6}\,\log(\mu)}{(|x_{12}||x_{13}||x_{23}|)^{2}}\,.
\end{equation}

Combining (\ref{leading order phi phi sigma}), (\ref{vertex corrections to phi phi sigma}),
(\ref{leading order phi phi sigma with dressed sigma legs}) and using the Callan-Symanzik equation\footnote{
There is no contribution from $\beta_h\,\frac{\partial}{\partial h}$ at the considered order ${\cal O}(\epsilon_{1,2}^{3/2})$.}
\begin{equation}
\left(\mu\,\frac{\partial}{\partial\mu} +\beta_g\,\frac{\partial}{\partial g} +\gamma_\sigma\right)\,\langle
\phi\phi\sigma\rangle = 0\,,
\end{equation}
we obtain
\begin{equation}
\label{beta g}
\beta_g\equiv \mu\,\frac{\partial}{\partial\mu}\,g = -\alpha\,\epsilon_1\,g+
\frac{g}{12(4\pi)^3}\,\left((N-12)\,g^2+h^2-12\,g\,h\right) +{\cal O}(\epsilon_{1,2}^2)\,.
\end{equation}
{}\\

Similarly, the three-point function $\langle \sigma\sigma\sigma\rangle$
at the tree level is given by
\begin{equation}
\langle \sigma (x_1)\sigma (x_2)\sigma (x_3)\rangle
=\frac{-h\,\mu^{\epsilon_1/2}\,C_\sigma^3\, U(2,2,2)|_{d=6}}{(|x_{12}||x_{13}||x_{23}|)^{2}}\,,
\end{equation}
while the linear order corrections are obtained by dressing the $\sigma$
legs of the leading order tree graph, and including the one-loop vertex corrections
\begin{center}
  \begin{picture}(163,114) (107,-10)
    \SetWidth{1.0}
    \SetColor{Black}
    \Line[](81,95)(81,51)
    \Line[](50,9)(110,10)
    \Line[](9,-7)(51,9)
    \Line[](151,-7)(110,10)
    \Line[](48,8)(81,51)
    \Line[](81,51)(110,10)
    \Vertex(50,9){4}
    \Vertex(81,51){4}
    \Vertex(110,10){4}
    \Vertex(10,-7){2}
    \Vertex(150,-7){2}
    \Vertex(81,95){2}
    \Text(87,70)[lb]{\scalebox{0.8}{$2\Delta_\sigma$}}
    \Text(45,30)[lb]{\scalebox{0.8}{$2\Delta_\phi$}}
    \Text(100,30)[lb]{\scalebox{0.8}{$2\Delta_\phi$}}
    \Text(74,-4)[lb]{\scalebox{0.8}{$2\Delta_\phi$}}
    \Text(17,5)[lb]{\scalebox{0.8}{$2\Delta_\sigma$}}
    \Text(130,5)[lb]{\scalebox{0.8}{$2\Delta_\sigma$}}
    \Text(180,30)[lb]{\scalebox{0.8}{$+$}}
    \Line[](281,95)(281,51)
    \Line[](250,9)(310,10)
    \Line[](209,-7)(251,9)
    \Line[](351,-7)(310,10)
    \Line[](248,8)(281,51)
    \Line[](281,51)(310,10)
    \Vertex(250,9){4}
    \Vertex(281,51){4}
    \Vertex(310,10){4}
    \Vertex(210,-7){2}
    \Vertex(350,-7){2}
    \Vertex(281,95){2}
    \Text(287,70)[lb]{\scalebox{0.8}{$2\Delta_\sigma$}}
    \Text(245,30)[lb]{\scalebox{0.8}{$2\Delta_\sigma$}}
    \Text(300,30)[lb]{\scalebox{0.8}{$2\Delta_\sigma$}}
    \Text(274,-4)[lb]{\scalebox{0.8}{$2\Delta_\sigma$}}
    \Text(217,5)[lb]{\scalebox{0.8}{$2\Delta_\sigma$}}
    \Text(330,5)[lb]{\scalebox{0.8}{$2\Delta_\sigma$}}
  \end{picture}
\end{center}
Carrying out the calculation, we arrive at
\begin{equation}
\label{beta h}
\beta_h\equiv \mu\,\frac{\partial}{\partial\mu}\,h = -\frac{h\,\epsilon_1}{2}+
\frac{1}{4(4\pi)^3}\,\left(-3\,h^3+N\,g^2\,(h-4\,g)\right) +{\cal O}(\epsilon_{1,2}^2)\,.
\end{equation}

\section{Fixed points}
\label{sec:fixed points}

We now proceed to studying fixed points of the beta-functions
$\beta_{g,h}$, given by (\ref{beta g}), (\ref{beta h}).
For general parameters $N$, $\alpha$, it is hard to find solutions to the equations $\beta_{g,h} = 0$
analytically. To circumvent this difficulty, we will begin by considering large $N$ limit, and
employ an algebraic $1/N$ expansion, given by (\ref{1/N expansion of couplings}).
Besides being a helpful analytic tool, in this case the algebraic $1/N$ expansion
will also conveniently allow us to compare the resulting
CFT data with the CFT data derived by using the perturbative diagrammatic $1/N$
expansion in the long-range $O(N)$ vector model (\ref{long range v.m. in HS}).

\begin{figure}[]
\begin{center}
\includegraphics[width=200pt]{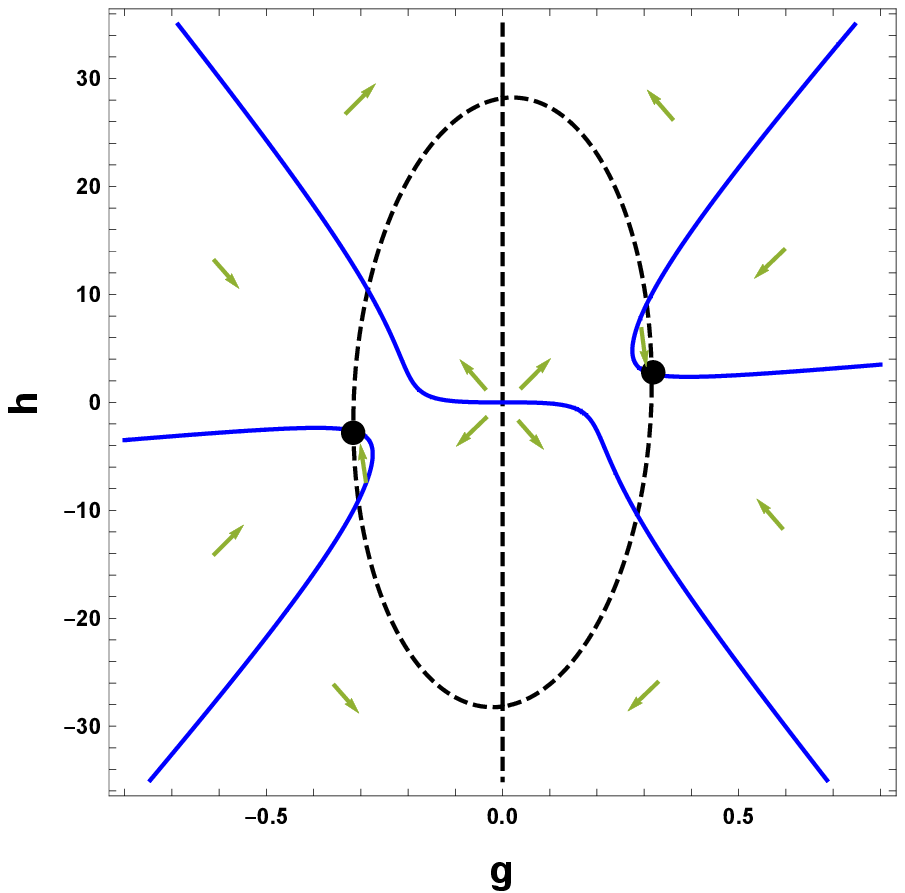}
\includegraphics[width=200pt]{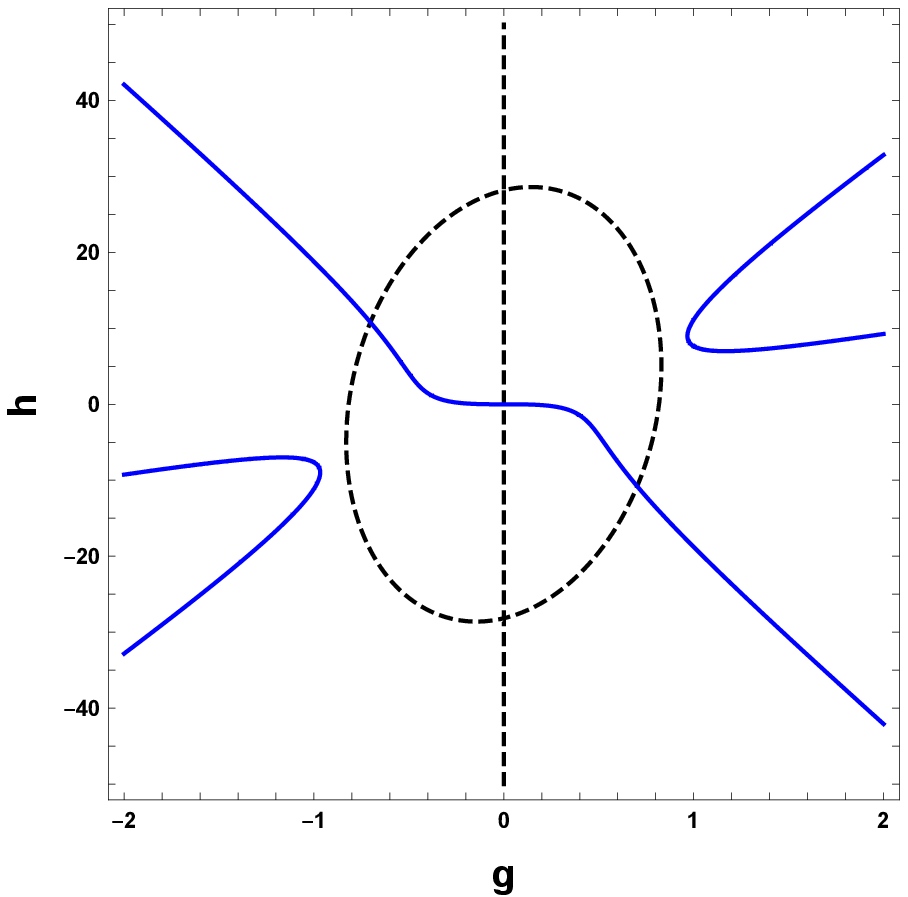}
\end{center}
\caption{Contour plots for $\beta_g$ (black, dashed) and $\beta_h$ (blue, solid) for $\epsilon_1=0.1$, $\alpha=1/3$, 
$N=8000$ (left), $N=1200$ (right). RG flow direction (towards IR), determined by the signs of $\beta_{g,h}$
in various domains of the plot, are indicated with arrows. When $N=8000$ (left) the RG flow has a pair of IR stable fixed
points on the $(g,h)$ plane (indicated with solid black dots). This fixed point for $\alpha=1/3$ disappears for
$N\leq N_{\textrm{crit}}(\alpha=1/3) =  2547$.
In particular, it is absent for $N=1200$ (right).}
\label{fig:betagh}
\end{figure}

The model (\ref{starting bilocal phi action}) admits pairs of physically equivalent fixed points
due to the $\mathbb{Z}_2$ symmetry,
\begin{equation}
\label{Z2 symmetry}
g\rightarrow -g\,,\quad  h\rightarrow -h\,,\quad \sigma\rightarrow -\sigma\,.
\end{equation}
Carrying out $1/N$ expansion of the $\beta_{g,h} = 0$ equations,
we obtain the following non-trivial solution\footnote{This procedure can be easily continued
to higher orders in $1/N$.}
\begin{equation}
\label{1/n solution to beta = 0}
\begin{aligned}
g_0 &= 16\,\sqrt{\frac{3\pi^3\alpha\epsilon_1}{N}}\,,\qquad\qquad
&&h_0 = \frac{24\alpha}{6\alpha-1}\,g_0\,,\\
g_1 &=\frac{6(1-36\alpha+132\alpha^2)}{(6\alpha-1)^2}\,,
&&h_1 = \frac{36\alpha(19+6\alpha(70\alpha-17))-18}{(6\alpha-1)^3}\,.
\end{aligned}
\end{equation}
The fixed point (\ref{1/n solution to beta = 0}) is in fact paired up with a physically equivalent fixed point,
obtained by the transformation $g_0\rightarrow -g_0$, $h_0\rightarrow -h_0$,
due to the symmetry (\ref{Z2 symmetry}).
Calculating determinant of the matrix of first derivatives of the beta functions
for the solution (\ref{1/n solution to beta = 0}), we get
\begin{equation}
{\cal M} \equiv \det\left(\frac{\partial (\beta_g,\beta_h)}{\partial(g,h)}\right) = \alpha(6\alpha-1)\epsilon_1^2
+{\cal O}\left(\frac{1}{N}\right)\,.
\end{equation}
Therefore ${\cal M} > 0$ in the region (\ref{bounds on alpha}),
indicating that the fixed point (\ref{1/n solution to beta = 0}) is IR stable on the $(g,h)$ plane.
The main statement of our paper is that the fixed point (\ref{1/n solution to beta = 0}) is described
by a CFT that is equivalent to the CFT in the UV critical regime of the non-local $O(N)$
vector model (\ref{long range v.m. in HS}). While we support this claim by a perturbative
calculation, we also suggest that such a duality holds true for the entire range (\ref{d/3 bound on s}).\\

\begin{figure}[]
\begin{center}
\includegraphics[width=300pt]{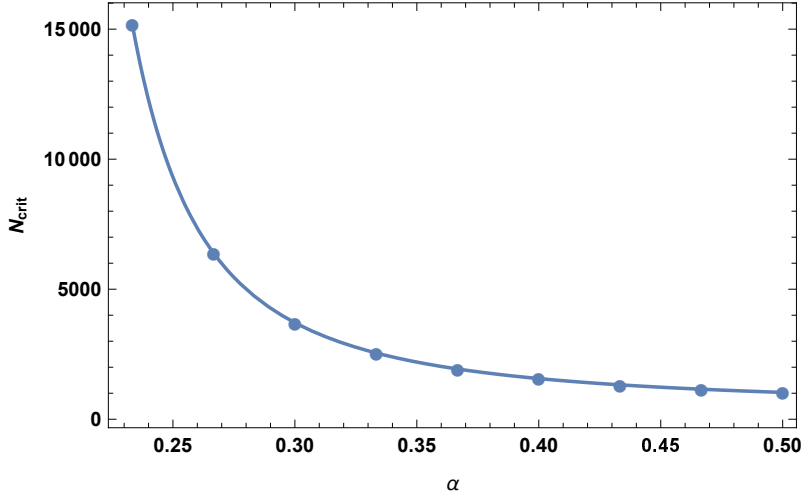}
\end{center}
\caption{$N_{\textrm{crit}}$ as a function of $\alpha$. An IR stable fixed
point on the $(g,h)$ plane exists only for large enough rank of the $O(N)$ group, $N\geq N_{\textrm{crit}}$.}
\label{fig:Nc_alpha}
\end{figure}

Before proceeding with further study of the CFT at the fixed point (\ref{1/n solution to beta = 0}),
let us discuss behavior of all of the fixed points of the beta functions $\beta_{g,h}$ at various values of $N$. 
The system of two cubic equations $\beta_{g,h} = 0$ can have up to nine different solutions.
One of the solutions represents a free UV fixed point $g=h=0$,
while eight other solutions are paired up due to the $\mathbb{Z}_2$ symmetry (\ref{Z2 symmetry}).
One pair of solutions is given by $g=0$, $h=\pm 8\pi ^{3/2}i\sqrt{2\epsilon_1/3}$,
and is not interesting for our purposes.

The remaining three pairs of solutions are accessible for general values of $\alpha$
and $N$ via numerics.
We observe that for general $\alpha$ and small $N$, defined by $N<N_{\textrm{crit}}(\alpha)$,
one has two complex-valued pairs of solutions, and a pair of real-valued solutions. The latter, however,
are not IR stable on the $(g,h)$ plane, and therefore cannot be connected to the large-$N$
solution (\ref{1/n solution to beta = 0}). However, for $N\geq N_{\textrm{crit}}(\alpha)$,
one of the pairs of the complex-valued solutions turns real-valued and satisfies
the requirement of the IR stability on the $(g,h)$ plane.
We illustrate such a behavior in figure~\ref{fig:betagh}, where we set $\epsilon_1 = 0.1$, $\alpha=1/3$,
and consider the cases of $N=8000$, $N=1200$.
In this case, $N_{\textrm{crit}}(\alpha=1/3) = 2548$. Correspondingly, one can see that the IR stable fixed point
exists on the $N=8000$ plot, but disappears on the $N=1200$ plot.

The value $N_{\textrm{crit}}(\alpha)$ can be determined analytically as a function of $\alpha$.
First, redefining $g = g_0\,x\,\sqrt{\epsilon_1}$, $h=h_0\,y\,\sqrt{\epsilon_1}$ we
can rewrite $\beta_{g,h} = 0$ equations as follows
\begin{equation}
\begin{aligned}
(N-12)\,g_0^2\, x^2-12\,g_0\,h_0\,x\,y+h_0^2 \, y^2-768\pi^3\,\alpha &=0\,,\\
4N\,g_0^3\,x^3-N\,g_0^2\,h_0\,x^2\,y+3h_0^3\,y^3+128\pi^3\,h_0\,y&=0\,.
\end{aligned}
\end{equation}
Written in such a form, this system of equations is in fact amenable to analytic solution.\footnote{This
can be done in \textit{Mathematica}.} For our purposes it is sufficient to focus on the behavior
of the square root term in one of the solutions, and determine the critical value of
$N$ above which it becomes real-valued. This yields
\begin{align}
k_1&=(-(1-6 \alpha )^6 \alpha  (18 \alpha +1)^2 (3 \alpha  (324 \alpha  (3 \alpha +1)+11)+2)^3 (12 \alpha  (\alpha  (72 \alpha -11)+3)-1)^\frac{1}{2}\notag\\
k_2&=(18 \alpha  (3 \alpha  (4 \alpha  (18 \alpha  (12 \alpha  (\alpha  (972 \alpha  (\alpha  (54 \alpha  (9 \alpha  (108 \alpha  (18 \alpha -1)+109)+43)+821)+93)\notag\\
&+7391)+381)+65)+215)-9)-5)+6 k_1-1)^\frac{1}{3}\label{Ncrit}\\
N_{\textrm{crit}}&=\frac{12}{(6 \alpha -1)^3 (18 \alpha +1)}\left(
34992 \alpha ^4-648 \alpha ^3+684 \alpha ^2+42 \alpha +k_2-2\right.\notag\\
&+\left.\frac{1}{k_2}\,
\left(108 \alpha ^2+1\right) (36 \alpha  (3 \alpha  (4 \alpha  (27 \alpha +2) (108 \alpha  (9 \alpha -1)+35)+3)-1)+1)\right)\,.\notag
\end{align}
Notice that $N_{\textrm{crit}}$ diverges when $\alpha\rightarrow 1/6$, i.e., at the lower
end of the window (\ref{bounds on alpha}).
We plot it in figure~\ref{fig:Nc_alpha}, where we 
 have also verified (\ref{Ncrit}) by testing it numerically for various values of $\alpha$.

\section{CFT data and critical duality}
\label{sec:CFT data}

In section~\ref{sec:fixed points} we discussed the structure of perturbative fixed points
of the couplings $g$, $h$ of the model (\ref{starting bilocal phi action}) near $d=6$ and $s=d/2-1$.
We found numerically that for each $\alpha$ in (\ref{bounds on alpha}) and large enough $N\geq N_{\textrm{crit}}(\alpha)$
there exists a real-valued fixed point that is IR stable on the $(g,h)$ plane. By expanding in $1/N$
in the large $N$ limit, we arrived at an analytical solution (\ref{1/n solution to beta = 0}) for the
coupling constants $g$, $h$ at this fixed point.

While the fixed point (\ref{1/n solution to beta = 0}) was determined perturbatively near $d=6$ and $s=d/2-1$,
we suggest that such an IR stable fixed point in fact exists for the range $d/3<s<\min(d/2, s_\star)$,
and that it is described by a CFT
that is equivalent to the critical UV regime of the 
long-range $O(N)$ vector model (\ref{long range v.m. in HS}). 
As we discussed in section~\ref{sec:intro}, one can see that such a critical
equivalence is indicated by the $1/N$ structure (\ref{gamma sigma from CS})
of the IR scaling dimension $[\sigma]_{\textrm{IR}}$ of the field $\sigma$, that can be easily
determined from the Callan-Symanzik equation (\ref{CS derivation of gamma sigma})
in the large $N$ limit.

Furthermore, from (\ref{gamma sigma}),  (\ref{1/n solution to beta = 0})
one obtains
\begin{equation}
\label{sigma IR dimension}
[\sigma]_{\textrm{IR}} = s + \frac{\epsilon_1}{N}\,\frac{12\alpha(30\alpha-1)}{6\alpha-1}
+{\cal O}\left(\epsilon_1^2,\frac{1}{N^2}\right)\,.
\end{equation}
On the other hand, expanding the anomalous dimension $\tilde\gamma_\sigma$
of the Hubbard-Stratonovich field (\ref{gamma sigma in v.m.}) in the long-range
$O(N)$ vector model in $\epsilon_{1}$ we obtain
\begin{equation}
\tilde \gamma_\sigma =\frac{\epsilon_1}{N}\,\frac{12\alpha(30\alpha-1)}{6\alpha-1}
+{\cal O}\left(\epsilon_1^2,\frac{1}{N^2}\right)\,,
\end{equation}
that is precisely in agreement with (\ref{sigma IR dimension}).

In a similar spirit, we can continue matching CFT data of the fixed point (\ref{1/n solution to beta = 0})
of the model (\ref{starting bilocal phi action})
with the critical long-range $O(N)$ vector model (\ref{long range v.m. in HS}).
One of the non-trivial consistency checks of the proposed critical
duality involves matching of the relative next-to-leading over leading in $1/N$ contributions to the
OPE coefficient of the three-point function $\langle\phi\phi\sigma\rangle$.
While at the fixed point (\ref{1/n solution to beta = 0}) 
of the model (\ref{starting bilocal phi action}) it is simply determined from
the tree level diagram due to the $g$ interaction vertex,
\begin{equation}
\label{phi phi sigma subleading over leading in cubic}
\langle\phi\phi\sigma\rangle\Bigg|_{\textrm{normalized}}=-\frac{C_\phi\,C_\sigma^\frac{1}{2}\,U(2,2,2)|_{d=6}\,g_0\,
\left(1+g_1/N + {\cal O}(\epsilon_1)\right)}{(|x_{12}||x_{13}||x_{23}|)^2}\,.
\end{equation}
in the long-range $O(N)$ vector model it was derived in \cite{Chai:2021arp} for general $d$, $s$, and
is given by (\ref{delta C phi phi sigma in v.m.}). Expanding the latter near $s=d/2-1$, $d=6$,
one obtains
\begin{equation}
\delta C_{\phi\phi\sigma}\Bigg|_{\textrm{normalized}}=\frac{1}{N}\frac{6(1-36\alpha+132\alpha^2)}{(6\alpha-1)^2}
+{\cal O}\left(\epsilon_1,\frac{1}{N^2}\right)\,,
\end{equation}
in agreement with $g_1/N$ due to (\ref{phi phi sigma subleading over leading in cubic}),
(\ref{1/n solution to beta = 0}).\footnote{
Importance of matching full CFT data, including sub-leading contributions to the OPE
coefficients, in the context of testing critical duality of vector models was recently
emphasized in \cite{Goykhman:2019kcj}.}

Continuing with the match of CFT data, we can perform $\epsilon_{1}$-expansion of the large-$N$ ratio
of the normalized OPE coefficients
$\langle\sigma\sigma\sigma\rangle/\langle\phi\phi\sigma\rangle$
in the model (\ref{long range v.m. in HS}),
given by (\ref{sigma sigma sigma over phi phi sigma in v.m.}), obtaining
\begin{equation}
\frac{\langle \sigma\sigma\sigma\rangle}{\langle \sigma\phi\phi\rangle} \Bigg|_{\textrm{normalized}}= 
\frac{24\alpha}{6\alpha-1}+{\cal O}\left(\epsilon_1,\frac{1}{N}\right)\,.
\end{equation}
Its counterpart in the CFT data at the fixed point (\ref{1/n solution to beta = 0}) 
of the model (\ref{starting bilocal phi action}) is given by the ratio $h_0/g_0$,
that it in fact agrees with.

\section{Discussion}
\label{sec:discussion}

In this paper we discussed the UV critical regime of the long-range $O(N)$ vector model
with quartic interaction (\ref{long range v.m. in HS}). Working at the one-loop order in perturbation
theory near $s=d/2-1$, $d=6$, we provided evidence for the critical duality
of this model and the IR fixed point of the cubic model coupled to a generalized free field (\ref{starting bilocal phi action}).
We suggest that such a critical universality in facts holds for the entire range of $d/3<s<\min(d/2, s_\star)$.
Additional checks can be performed by matching more CFT data
on both sides of the proposed duality, in particular by combining the techniques of $\epsilon$-
and $1/N$-expansion, and studying higher orders thereof.

It would be interesting to see whether the model (\ref{starting bilocal phi action})
at criticality can be obtained by deforming the critical cubic model of  \cite{Fei:2014yja}
by coupling it to a generalized free field $\chi$ of dimension $[\chi] = d - [\phi] = (d+s)/2$,
analogously to the case of the long-range $\phi^4$ model considered in \cite{Behan:2017dwr,Behan:2017emf}.

It was recently shown that critical long-range vector models play
an important role in systems exhibiting the phenomenon of persistent symmetry breaking \cite{Chai:2021djc}.\footnote{
See \cite{Chai:2020zgq,Chai:2020hnu} for other recent work on persistent symmetry
breaking in vector models.} The cubic model proposed in this paper can shed additional
light on properties of the corresponding CFTs.

\section*{Acknowledgements} \noindent  
The work of S.C. is supported by the Infosys Endowment for the study of the Quantum Structure of Spacetime.
The work of M.G. is partially supported by the Binational Science Foundation (grant No. 2016186), the Israeli Science Foundation Center of Excellence (grant No. 2289/18), and by the Quantum Universe I-CORE program of the Israel Planning and Budgeting Committee (grant No. 1937/12). 

\appendix

\section{Some useful identities}
\label{app_a}

In this appendix we collect some known expressions and identities, that are useful
to carry out perturbation theory calculations in position space.

Loop diagram in position space are simply additive:
\begin{center}
  \begin{picture}(257,50) (0,0)
    \SetWidth{1.0}
    \SetColor{Black}
    \Arc[clock](81,-39)(77.006,127.614,52.38600001)
    \Arc[](80,86)(80,-126.87,-53.13)
    \Line[](160,22)(256,22)
    \Vertex(160,22){2.001}
    \Vertex(256,22){2.001}
    \Vertex(33,22){2.001}
    \Vertex(129,22){2.001}
    \Text(142,20)[lb]{$=$}
    \Text(77,-5)[lb]{\scalebox{0.8}{$2b$}}
    \Text(77,43)[lb]{\scalebox{0.8}{$2a$}}
    \Text(190,27)[lb]{\scalebox{0.8}{$2(a+b)$}}
  \end{picture}
\end{center}

The propagator merging relation identity is given by
\begin{equation}
\label{prop merging}
\int d^d x_2\, \frac{1}{|x_2|^{2a}|x_1-x_2|^{2b}}
=U(a,b,d-a-b)\,\frac{1}{|x_{1}|^{2a+2b-d}}\,,
\end{equation}
where we defined
\begin{align}
\label{U def}
U(a,b,c) &= \pi^\frac{d}{2} A(a)A(b)A(c)\,.
\end{align}
Here we have introduced
\begin{equation}
A(x) = \frac{\Gamma\left(\frac{d}{2}-x\right)}{\Gamma(x)}\,.
\end{equation}
This relation can also be represented diagrammatically as
\begin{center}
  \begin{picture}(98,10) (130,-60)
    \SetWidth{1.0}
    \SetColor{Black}
    \Line[](30,-58)(90,-58)
    \Line[](90,-58)(150,-58)
    \Line[](180,-58)(240,-58)
    \Vertex(30,-58){2.0001}
    \Vertex(90,-58){4.0001}
    \Vertex(150,-58){2.0001}
    \Vertex(180,-58){2.0001}
    \Vertex(240,-58){2.0001}
    \Text(55,-53)[lb]{\scalebox{0.801}{$2a$}}
    \Text(115,-53)[lb]{\scalebox{0.801}{$2b$}}
    \Text(163,-61)[lb]{$=$}
    \Text(180,-53)[lb]{\scalebox{0.801}{$2(a+b)-d$}}
    \Text(250,-63)[lb]{$\times U(a,b,d-a-b)$}
  \end{picture}
\end{center}

Uniqueness relation, valid for $a_1+a_2+a_3=d$, has the form \cite{DEramo:1971hnd,Symanzik:1972wj}
\begin{equation}
\label{uniqueness}
\int d^dx\,\frac{1}{|x_1-x|^{2a_1} |x_2-x|^{2a_2} |x_3-x|^{2a_3}} 
=\frac{U(a_1,a_2,a_3)}{|x_{12}|^{d-2a_3}|x_{13}|^{d-2a_2}|x_{23}|^{d-2a_1}}\,,
\end{equation}
and can be represented graphically as
\begin{center}
  \begin{picture}(210,66) (70,-31)
    \SetWidth{1.0}
    \SetColor{Black}
    \Line[](32,34)(80,2)
    \Line[](80,2)(32,-30)
    \Line[](80,2)(128,2)
    \Line[](192,34)(192,-30)
    \Line[](192,-30)(240,2)
    \Line[](240,2)(192,34)
    \Vertex(32,34){2.001}
    \Vertex(32,-30){2.001}
    \Vertex(80,2){4.001}
    \Vertex(128,2){2.001}
    \Vertex(192,34){2.001}
    \Vertex(192,-30){2.001}
    \Vertex(240,2){2.001}
    \Text(55,-25)[lb]{\scalebox{0.801}{$2a_1$}}
    \Text(55,22)[lb]{\scalebox{0.801}{$2a_2$}}
    \Text(100,5)[lb]{\scalebox{0.801}{$2a_3$}}
    \Text(160,-1)[lb]{$=$}
    \Text(180,-1)[lb]{\scalebox{0.801}{$\alpha$}}
    \Text(215,-28)[lb]{\scalebox{0.801}{$\beta$}}
    \Text(215,24)[lb]{\scalebox{0.801}{$\gamma$}}
    \Text(250,-5)[lb]{$\times \left(-\frac{2}{\sqrt{N}}\right) U\left(a_1,a_2,a_3\right)$}
  \end{picture}
\end{center}
Here we have defined $\alpha = d-2a_3$, $\beta = d-2a_2$, $\gamma = d-2a_1$.

\end{document}